%Paper: 9110014
%From: PLESSER%YALPH2.BITNET@YALEVM.YCC.Yale.Edu
%Date: Fri, 4 Oct 1991 01:44 EDT

\input harvmac.tex
\nopagenumbers

\Title{\vbox{\baselineskip12pt\hbox{CLNS 91-1109}\hbox{YCTP-P32-91}}}
{\vbox{\centerline{Mirror Manifolds: A Brief Review and Progress Report}}}
\footnote{ }{Based in part on invited talk presented at the second
International PASCOS conference, Northeastern University, 1991.}

\centerline{B.R. Greene }
\medskip\centerline{F.R. Newman Laboratory of Nuclear Studies}
\centerline{Cornell University}\centerline{Ithaca, NY \ 14853}
\vskip.2in
\centerline{M.R. Plesser }
\medskip\centerline{Department of Physics}
\centerline{Yale University}\centerline{New Haven, CT \ 06511}
\vskip .2in
We first give a complete, albeit brief, review of the discovery of mirror
symmetry in $N=2$ string/conformal field theory. In particular, we
describe the naturality arguments which led to the initial mirror
symmetry conjectures and the subsequent work which established the
existence of mirror symmetry through direct construction. We then
review a number of striking consequences of mirror symmetry -- both
conceptual and calculational.  Finally, we describe recent work which
introduces a variant on our original proof of the existence of mirror
symmetry. This work affirms classical--quantum symmetry duality as well as
extends the domain of our  initial mirror symmetry construction.

\Date{9/91}

\noblackbox
\font\bigrm=cmr10 scaled \magstephalf
\def\inbar{\,\vrule height1.5ex width.4pt depth0pt}
\font\cmss=cmss10 \font\cmsss=cmss10 at 7pt
\def\BZ{\relax\ifmmode\mathchoice
{\hbox{\cmss Z\kern-.4em Z}}{\hbox{\cmss Z\kern-.4em Z}}
{\lower.9pt\hbox{\cmsss Z\kern-.36em Z}}
{\lower1.2pt\hbox{\cmsss Z\kern-.36em Z}}\else{\cmss Z\kern-.4em Z}\fi}
\def\IC{\relax\hbox{$\inbar\kern-.3em{\rm C}$}}
\def\IP{\relax{\rm I\kern-.18em P}}

\def\Tr#1{\hbox{{\bigrm Tr}\kern-1.05em \lower2.1ex \hbox{$\scriptstyle#1$}}\,}
\def\s{super}

\def\27b{\overline {27} }
\def\b27{\overline {27} }

\def\ex#1{\hbox{$\> e^{#1}\>$}}
\def\CP#1{\hbox{$\hbox{\IC \IP}^{#1}$}}
\def\WCP#1#2{\hbox{$\hbox{W\IC \IP}^{#1}_{#2}$}}
\def \ie{\hbox{\it i.e.}\ }

\def\cft{conformal field theory}

\def\CY{Calabi-Yau }

\nref\rRonover
{M.B. Green, J.H. Schwarz, and L. Brink, Nucl. Phys. {\bf B198} (1982) 474
\semi
K. Kikkawa and M. Yamasaki, Phys. Lett. {\bf B149} (1984) 357 \semi
N. Sakai and I. Senda, Prog. Theor. Phys. {\bf 75} (1984) 692.}%
\nref\rDVV{R.Dijkgraaf, E. Verlinde and H. Verlinde, Comm. Math. Phys.
{\bf 115} (1988) 649.}%
\nref\rLVW{W. Lerche, C. Vafa, and N.P. Warner, Nucl. Phys. {\bf B324} (1989)
427.}%
\nref\rDIXON{L. Dixon, V. Kaplunovsky, and J. Louis, Nucl. Phys. {\bf B329}
(1990) 27.}
\nref\rGP{B.R. Greene and M.R. Plesser, Nucl. Phys. {\bf B338} (1990) 15.}%
\nref\rCLS{P. Candelas, M. Lynker, and R. Schimmrigk, Nucl. Phys. {\bf B341}
(1990) 383.}%
\nref\rBGWP{B. Greene, Comm. Math. Phys. {\bf 130} (1990) 335.}%
\nref\rCGOP{P. Candelas, X.C. de la Ossa, P.S. Green, and L. Parkes,
Nucl. Phys. {\bf B359} (1991) 21; Phys. Lett. {\bf 258B} (1991) 118.}%
\nref\rALR{P.S. Aspinwall, C.A. L\"utken, and G.G. Ross, Phys. Lett. {\bf 241B}
(1990) 373.}%
\nref\rAL{P.S. Aspinwall and C.A. L\"utken, Nucl. Phys. {\bf B355}
(1991) 482.}%
\nref\rGEP{D. Gepner, Phys. Lett. {\bf 199B} (1987) 380;
Nucl. Phys. {\bf B296} (1987) 380. }%
\nref \rGVW{B.R. Greene, C. Vafa and N.P. Warner,
Nucl. Phys. {\bf B324} (1989), 371.}%
\nref\rDG{J. Distler and B.R. Greene, Nucl. Phys.
{\bf B309} (1988) 295.}%
\nref\rTONE{A. Strominger, Phys. Rev. Lett. {\bf 55} (1985) 2547.}%
\nref\rANS{A. Strominger and E. Witten, Comm. Math. Phys. {\bf 101} (1985)
341.}%
\nref\rDSWW{M. Dine, N. Seiberg, X. G. Wen, and E. Witten, Nucl. Phys.
{\bf B278} (1987) 769; {\bf B289} (1987) 319.}%
\nref\rdmrsn{D. Morrison, ``Mirror Symmetry and Rational Curves on Quintic
Threefolds: A Guide for Mathematicians'', Duke Preprint DUK-M-91-01.}%
\nref\rqs{C.Vafa, Mod. Phys. Lett. A4; 1169, 1989; Mod. Phys. Lett. A4,
1615, 1989.}%
\nref\rLS{R. Schimmrigk and M. Lynker, Phys. Lett. {\bf B249} (1990) 237.}%

\newsec{Introduction}

Classical solutions for string theory are described by conformal field
theories. The most interesting solutions, for fundamental as well as
``phenomenological'' reasons, are those which yield spacetime
supersymmetric theories: $N=2$ superconformal models.
Superconformal field theories which arise from sigma
models on appropriately chosen target spaces
comprise an especially interesting category of vacuum solutions.
These solutions which admit a geometrical interpretation are extremely useful
probes of the unusual and sometimes remarkable properties of string theory.
The reason for this is clear: with a geometrical interpretation in hand,
we can perform direct comparisons between the string description and more
conventional analyses based on general relativity and/or quantum mechanics.

A striking feature of string theory which emerges from this analysis is the
existence of {\it duality} transformations. These transformations are
symmetries
of string physics which, when interpreted as geometrical operations, relate
distinct geometrical configurations.
The simplest example of this notion is provided by $c=1$ string theory.
With the geometrical interpretation of this solution as
string propagation on a circle, one finds
the surprising conclusion  that physics is invariant under $R \rightarrow
1/2R$ where $R$ is the radius of the circle \rRonover.
This phenomenon has no analog in conventional
point particle quantum mechanics
nor in general relativity in such a context.
The physical interpretation of this symmetry is related to
the existence of ``winding modes'' in which the string wraps around the
spacetime; this feature of string physics may be generalized to toroidal
spacetimes and is an important signature of truly stringy physics.

Toroidal spacetimes are in fact a small subset of a class of classical string
vacua for which a geometrical interpretation is known -- the class of \CY\
manifolds.
Mirror
symmetry provides an extremely robust generalization of such ideas
into this far larger class of compactifications. It
identifies {\it topologically distinct} string backgrounds which give
rise to {\it identical } physics.
By providing a link between {\it a priori} unrelated manifolds, mirror
symmetry establishes  a powerful conceptual and calculational tool from
the points of view of both physics and mathematics.
Although remarkable from the mathematical vantage point (and quite
powerful, as we shall discuss) the existence of mirror manifolds is
rather natural from the point of view of physics. In fact, several
groups, based on `naturality' arguments, had earlier speculated on the
possibility of mirror symmetry.
Let us now turn to the motivation for
such speculation.

There are two types of moduli on a Calabi-Yau manifold%
\foot{We emphasize Calabi-Yau threefolds for most of our
discussion as this is the most natural dimension from the
string theoretic point of view. Essentially all of our
results, however, immediately extend to higher dimension.}%
: those associated
with the complex structure and those associated with the K\"ahler structure.
The former are described by the cohomology group $H^{2,1}$ while the
latter by $H^{1,1}$%
\foot{We recall that when describing string propagation we are naturally led to
a complexification of the K\"ahler cone when we augment the metric with the
background antisymmetric tensor field $B$.}.
{}From the mathematical point of view these are
vastly different objects. Intuitively, the complex structure moduli
parametrize deformations of
the shape of the Calabi-Yau manifold while the K\"ahler moduli
determine its size.
These deformations of the manifold are described by
deformations of the associated \cft\ by
marginal operators. Like the deformations of the related geometrical structure,
these come in two varieties: those with $(\Delta = \overline
\Delta = 1/2; \quad Q = \overline Q = 1)$ and those with
$(\Delta = \overline
\Delta = 1/2; \quad Q = -\overline Q = 1)$,
where $\Delta$ denotes the conformal
weight and $Q$ denotes the charge under the $U(1)$ subgroup of the $N=2$
superconformal algebra%
\foot{The quantum numbers given here refer
to the lower component of the supermultiplet to which the given operator
belongs.}.
Unlike their geometrical counterparts, the difference between these two kinds
of operators has no intrinsic significance.
The
sign of the left moving $U(1)$ charge is essentially a matter of
convention. When we identify this conformal field theory with string
propagation on a Calabi-Yau manifold, we must identify one of these two
sets of fields with elements in $H^{1,1}$ and the other set of fields with
elements in $H^{2,1}$. It seems somewhat strange that geometry so greatly
exaggerates the difference between these two kinds of conformal fields.

The above led Dixon
to speculate that a more natural state of affairs would ensue
if there were two geometrical interpretations of the conformal field theory
which differ by interchanging the identifications of conformal fields
with differential forms.
In this way, both {\it a priori} possible identifications of conformal fields
with differential forms would be realized.
The naturality of this speculation was elaborated
upon by Lerche, Vafa and Warner
\rLVW\ (see also \rDIXON)
who found that the marginal operators discussed above (or rather, their lower
components) are elements in two distinct rings which are present in any $N=2$
\s conformal theory. What these authors found is that in models with a
geometrical interpretation, one of these two rings is
isomorphic to a deformation of the cohomology ring of the underlying manifold.
The other ring has no obvious geometric interpretation
As pointed out in \rLVW, it seems strange that only one conformal field theory
ring has a geometrical interpretation. A more natural situation would again
follow if there were two Calabi-Yau manifold interpretations of such
a conformal field theory in which the identification of fields with forms
would be interchanged\foot{These two rings are shown in \rLVW\ to be
connected via spectral flow. Thus, another way of phrasing the speculation
in \rLVW\ is that {\it if} spectral flow
could be given a geometrical interpretation, then it would produce the
desired pair of Calabi-Yau spaces. As yet, no such geometrical interpretation
of spectral flow has been found.}.
Thus, each manifold would provide the geometrical
interpretation for one of the two conformal field theory rings.

These speculations based on naturality from the point of view of physics
should be juxtaposed with similar reasoning from the viewpoint of mathematics.
The  hypothesized pairing of manifolds described above
implies the existence of Calabi-Yau
manifolds whose $(2,1)$ and $(1,1)$ Hodge numbers are interchanged%
\foot{Of course, giving rise to the same conformal field theory
(the same physics) implies more than this, as we will
discuss.}.
For a variety of technical reasons
(including the supposed unlikely existence
of Calabi-Yau manifolds with $h^{1,1}$ as large as some of the known
values of $h^{2,1}$) it was felt by a number of mathematicians
\ref\rY{S.-T. Yau, private communication.}\
that it would be rather {\it unnatural} for such pairs to exist.

These conflicting intuitions were partially resolved by solid arguments in
\rGP\ in which the existence of conformal field theories with
two geometrical interpretations, in the sense described above, was
proven. For a particular class of conformal field
theories (those based on products of minimal models and deformations
thereof), we explicitly found the dual geometrical interpretations in terms
of manifolds differing by the interchange of $H^{1,1}$ and $H^{2,1}$.
Because this interchange corresponds to a reflection of the manifold's
Hodge diamond
(which is equivalent to a ninety--degree rotation of the
diamond), we gave the name {\it mirror manifolds}
\rGP\ to such
a pair of Calabi-Yau spaces.
We will review this work in the next section.

The arguments of \rGP\ are valid for a rather restricted class of \CY\
manifolds, but there is strong evidence that the existence of mirror manifolds
is not.
Simultaneous with the above work, Candelas, Lynker and Schimmrigk
\rCLS\ completed
a thorough computer search of Calabi-Yau manifolds in weighted projective
four space initiated in \rBGWP. They found that the set of all such
constructions is almost invariant under the interchange of $h^{1,1}$ and
$h^{2,1}$.
While finding two manifolds differing by the interchange of
these Hodge numbers by no means assures that the manifolds are a mirror pair
(to be a mirror pair they must correspond to the same conformal field theory)
this data indicates that mirror duality probably extends beyond
the particular class of models treated in \rGP.
We have subsequently shown
\ref\rGPF{B.R. Greene and M.R. Plesser, in preparation.}\
that at least some of the candidate mirror pairs
found in \rCLS\ do in fact correspond to isomorphic conformal field theories,
but there are
unexpected subtleties in the argument, which will be presented in section four.

The existence of the mirror manifold $M'$ implies some rather strong and
sometimes startling conclusions regarding the physics and mathematics of $M$.
Section three is devoted to a discussion of the most striking of these, and to
a number of recent papers
\refs{\rCGOP-\rAL}\
in which these results have been verified
and elaborated upon by various groups.
Section five offers some brief conclusions and mentions
some open issues in mirror symmetry.

\newsec{Constructing Mirror Manifolds}

The basic strategy utilized in \rGP\ for the construction of
mirror manifolds is straightforward to describe. We begin with
a class of conformal field theories whose geometrical interpretation
in terms of string propagation on Calabi-Yau spaces has been known
for some time
\refs{\rGEP,\rGVW}: tensor products of $N=2$ minimal
model conformal field theories
(with an appropriate projection onto states with integral
$U(1)$ charges \rGEP). These were shown in \rGVW\ to correspond
to Calabi-Yau hypersurfaces in weighted projective space.

Our strategy for building mirror manifolds is to seek
a nontrivial automorphism of the conformal field theory
which is {\it not} an automorphism of the geometrical
description. Rather, we hope that this sought for mapping will
yield an isomorphic conformal theory while producing a new corresponding
Calabi-Yau manifold, topologically distinct from the initial one.

How can we find such a mapping? Recall from our earlier discussion that
it is only the relative sign of the left moving $U(1)$ charge which
distinguishes amongst the two kinds of conformal field theory moduli.
Imagine the consequences of an operation  $\cal O$
on the conformal field theory $K$
which
generates an isomorphic theory with the explicit isomorphism requiring
flipping the sign of all left  moving $U(1)$ charges.
To do so, let us further imagine
that this operation descends to the geometrical description
and hence has a legitimate interpretation as an action on the initial
Calabi-Yau space, $M$. By legitimate here we mean, of course, that the range of
our operation is contained within the space of Calabi-Yau manifolds.
Now we can extract the consequences:
Both $M$ and ${\cal O}(M)$ correspond to the same conformal field theory since
$K$ and ${\cal O}(K)$ are isomorphic. Furthermore, since the explicit
isomorphis
   m
between $K$ and ${\cal O}(K)$ is the reversal of  the sign of the
left moving $U(1)$ charges,
the identification of conformal fields with differential forms is {\it
reversed}
on ${\cal O}(M)$ relative to $M$. Thus $M$ and ${\cal O}(M)$ would constitute
a mirror pair, if such an operation $\cal O$ could be found. In particular,
we would have
\eqna\eHOD
$$\eqalignno{h^{2,1}_M &= h^{1,1}_{{\cal O}(M)} & \eHOD a\cr
 h^{1,1}_M &= h^{2,1}_{{\cal O}(M)} \> .& \eHOD b\cr}$$

In \rGP\ such an operation $\cal O$ was constructed.
Its existence relies on the notion of `orbifolding'.
If a manifold $M$ is invariant under a group of symmetries
$G$ (in our case these will generally be holomorphic automorphisms)
we can consider the quotient space $M' = M/G$.
We restrict attention to groups $G$ which ensure that $M'$ is also Calabi-Yau.
Such $G$ are called `allowable' \foot{We note that $G$ is allowable if
it preserves the holomorphic $n$--form present on the initial Calabi-Yau
$n$--fold.}.
Similarly, if a conformal
theory $K$ respects a symmetry group $G$, we can consider the quotient theory
$K/G$.
The crucial property of these operations for our argument, is that the quotient
conformal theory describes propagation on the
quotient of the underlying Calabi-Yau manifold
by the same allowable group actions $G$%
\foot{To be more precise, the quotient will have singularities at the fixed
points of the $G$ action. The orbifold \cft\ corresponds to
a singular limit of an appropriate
desingularization of the quotient. The condition on $G$ described above assures
us of the existence of a desingularization which is itself a Calabi-Yau
manifold.}. The correspondence between group
actions in the two pictures is furnished by the identification of
homogeneous coordinates on the manifold with primary fields in the conformal
field theory given by the arguments of \rGVW.

We can now describe
the sought for operation: $\cal O$ is  taken to be orbifolding by the maximal
allowable $G$ which is a subgroup of the group of holomorphic scaling
symmetries. We will not reproduce the proof of this statement here, referring
the interested reader to \rGP\ for details, but will attempt to sketch the
salient points.
To begin, we leave geometry and focus
on a single $N=2$ minimal model at level $P$. Such a theory respects
a $\BZ_{P+2}$ scaling symmetry. Furthermore, the effect of orbifolding by
this $\BZ_{P+2}$ is to yield an isomorphic theory with the explicit isomorphism
consisting of the reversal of all left moving $U(1)$ charges
\ref\rGQ{D. Gepner and Z. Qiu, Nucl. Phys. {\bf B285} [FS19]
(1987) 423.}.
We can now apply this fact
to ($U(1)$ projected) tensor products of minimal models
at levels, say, $P_1,...,P_5$. (For more general cases see \rGP.)
We denote the $U(1)$ projected theory by $(P_1,...,P_5)$ and  we
recall \rGVW\ the conformal theory -- Calabi-Yau identification
\eqn\eID{(P_1,P_2,P_3,P_4,P_5) \rightarrow Z_1^{P_1+2} + Z_2^{P_2 +2} +
 Z_3^{P_3 +2} + Z_4^{P_4 +2} + Z_5^{P_5 +2} = 0}
where the right-hand side is a Calabi-Yau hypersurface in
weighted projective four space
$\WCP{4}{{d \over P_1 + 2}, {d \over P_2 + 2}, {d \over P_3 + 2},
{d \over P_4 + 2}, {d \over P_5 + 2} }$, where $d$ is the degree of homogeneity
of the defining polynomial.
Both sides respect a group of scaling symmetries%
\foot{This is in fact a bit naive. A $\BZ_d$ subgroup of this group acts
trivially on both sides; on the \cft\ side this is because we have already
projected to states invariant under this action to obtain a string vacuum,
while on the geometry side this is result of the fact that the $Z_i$ are
homogeneous coordinates on a projective space.}
$S = \BZ_{P_1 + 2}
\times \cdots \BZ_{P_5 +2}$. Let $G \subset S$ be the maximally allowable
subgroup of $S$. We claim that $K/G$ is isomorphic to $K$ under the
isomorphic mapping which flips the sign of all leftmoving $U(1)$ charges.
Furthermore, following our discussion above, $M' = M/G$ is a new
topologically distinct Calabi-Yau manifold which corresponds to the same
conformal theory; in particular $h^{2,1}_{M'} = h^{1,1}_M$ and
$h^{1,1}_{M'} = h^{2,1}_M$.
Our remarks in the previous paragraph regarding a single minimal model
hint at the underlying justification of our assertion. However, passing
from $S$ to $G$ and implementing the $U(1)$ projection are
additional complications over the single model case which must be dealt with.
In fact, each complication turns out to be crucial in resolving the other
\rGP.

To be concrete, let us recall the simplest example of this
mirror manifold construction.
Consider the theory $(3,3,3,3,3)$ which corresponds to the Fermat
quintic hypersurface in \CP4:
\eqn\eQ{(3,3,3,3,3) \rightarrow Z_1^5 + Z_2^5 + Z_3^5 + Z_4^5 + Z_5^5 = 0.}
It is immediate to check that $S = (\BZ_5)^5$ while $G = (\BZ_5)^3$. Thus,
the results of \rGP\ show that
\eqna\eQQ
$$\eqalignno{ &Z_1^5 + Z_2^5 + Z_3^5 + Z_4^5 + Z_5^5 = 0 &\eQQ a \cr
\noalign{\hbox{and}}
&{Z_1^5 + Z_2^5 + Z_3^5 + Z_4^5 + Z_5^5 =0 \over {(\BZ_5)^3} } &\eQQ b \cr}$$
are mirror manifolds\foot{The $(\BZ_5)^3$ action
is given in table 2.}.
Both of these Calabi-Yau spaces correspond to the conformal theory
$(3,3,3,3,3)$ and as we see from table 2, their Hodge numbers are appropriately
interchanged.
In table 1 we list some other examples.

Before discussing the implications of mirror symmetry, we pause here
to emphasize three important points. First, our construction works equally
well on any orbifold of the theories under discussion. That is, the
space of all orbifolds of a given theory can be partitioned into mirror
pairs. We illustrate this with the quintic hypersurface in table 2.
We note that the mirror of a theory $M/H$ with $H \subset G$ is given by
$M/H^*$ where $H^*$ is the complement of $H$ in $G$ (that is, the smallest
group containing $H$ and $H^*$ is $G$).
Second, our arguments are not specific to complex dimension three and
immediately generalize to other dimensions.
 Third, our discussion to this point
has been tied to very special points in the respective  Calabi-Yau moduli
spaces. Namely, we have focused on Fermat points as these correspond
to the well understood minimal model conformal field theories.
By deformation arguments we can immediately extend our results
to more general points in moduli space. For example, let $M$
(a Fermat hypersurface) and $M'$ be
mirror manifolds, each of which corresponds to the conformal field theory
$K$. Let us deform, say, the complex structure of $M$. This corresponds to
deforming $K$ by a particular and identifiable marginal operator.
On $M'$ this conformal field theory operator is interpretable as a K\"ahler
modulus, due to the flipped sign of the $U(1)$ charge, as discussed.
Hence, the mirror of the complex structure deformed $M$ is
this K\"ahler deformation of $M'$. Quite generally, then, we can follow this
procedure to determine the mirror of any deformation of $M$\foot{We should
acknowledge at this point the possibility of subtleties in
 performing this identification  globally in the respective moduli spaces.
Mathematically, then, it would be more precise to limit our analysis to local
deformations only although we believe our statements to be true globally as
well.}. Hence, modulo the caveat in the footnote, we have the
main result of \rGP:

{\it For any Calabi-Yau hypersurface $M$ which belongs to a moduli space
that admits a Fermat point, there exists a mirror manifold (in the full
sense of conformal field theory) $M'$. At the Fermat point, $M$ and $M'$ are
related by $M' = M/G$ where $G$ is the maximal allowable subgroup of the
group of scaling symmetries.}

We should note that even this statement is not the most general we could
make, since for example, as shown in \rGP\
and illustrated by the third line of table 1,
the hypersurface constraint
can be relaxed. As mentioned in the introduction,
we have reason to believe that even then
the class of models for which the above argument works is not the largest for
which mirror manifolds exist.
Having outlined the arguments by which mirror symmetry was shown to
exist, let us briefly turn to some consequences. In
\rGP\ we described in some detail a number of important implications
of mirror symmetry. In the next section we review the arguments that lead to
these claims, as well as some subsequent work by other groups which
lead to a rather spectacular verification of their veracity as well as to some
new applications.

\newsec{Applications of Mirror Symmetry}

Having  reviewed the initial speculations and subsequent work which
established the existence of mirror symmetry, we would now like to turn
to a discussion of the implications of this phenomenon, as well as
some recent work applying mirror symmetry to interesting
and explicit examples.

Let $M$ and $M'$ be mirror Calabi-Yau manifolds each corresponding to the
conformal field theory $K$. Consider a (non--vanishing) three point
function of \cft\ operators corresponding to
$(2,1)$ forms on $M$. Mathematically, this
correlation function is given by the simple integral on $M$ \rTONE\
\eqn\eCOUP{\int_{M} \omega^{abc}
\tilde b^{(i)}_a \wedge \tilde b^{(j)}_b \wedge \tilde b^{(k)}_c \wedge \omega}
where the $\tilde b^{(i)}_a$ are $(2,1)$ forms (expressed as elements
of $H^1(M',T)$ with their subscripts being tangent space indices)
and $\omega$ is the holomorphic three form.
Due to the nonrenormalization theorem proved in \rDG, we know that this
expression \eCOUP\ is the exact conformal field theory result.
By mirror symmetry, these same \cft\ operators correspond to particular and
identifiable $(1,1)$ forms on the mirror $M'$, which we can label $b^{(i)}$.
Mathematically, (due to the absence of
a nonrenormalization theorem) the expression for such a coupling \rANS,\rDSWW\
in terms of geometric quantities on $M'$ is comparatively complicated:
\eqn\eCOUPT{ \int_{M'} b^{(i)} \wedge b^{(j)} \wedge b^{(k)} +
\sum_n \ex{(-nR)}
\sum_{I_n}  ( \int_{I_n} X^*(b^{(i)} )) ( \int_{I_n} X^*(b^{(j)} ))
( \int_{I_n} X^*(b^{(k)} )) }
where $I_n$ is a holomorphic
instanton of charge $n$, $X$ is the map of the (worldsheet) instanton
into $M'$,
and $R$ is the radius of $M'$.
The first term in \eCOUPT\
is the topological triple intersection form on $M'$.

Now, since both \eCOUP\ and \eCOUPT\ correspond to the {\it same}
conformal field theory correlation function, they must be equal; hence we
have \rGP\
\eqna\eEQUAL
$$\displaylines{
\int_{M} \omega^{abc}
\tilde b^{(i)}_a \wedge \tilde b^{(j)}_b \wedge \tilde b^{(k)}_c
\wedge \omega  = \hfill\eEQUAL{} \cr
\int_{M'} b^{(i)} \wedge b^{(j)} \wedge b^{(k)} +
\sum_n \ex{(-nR)}
\sum_{I_n}  ( \int_{I_n} X^*(b^{(i)} )) ( \int_{I_n} X^*(b^{(j)} ))
( \int_{I_n} X^*(b^{(k)} ))\> .}$$
Notice the crucial role played by the underlying conformal field
theory in deriving this equation. If we simply had two manifolds whose
Hodge numbers were interchanged we could not, of course, make any
such statement.
This result \rGP\ is rather surprising and clearly very powerful.
We have related expressions on {\it a priori} unrelated manifolds
which probe rather intimately the structure of each.  Furthermore,
the left hand side of \eEQUAL{}\ is directly calculable
while the right hand side requires, among other things, knowledge
of the  rational curves of every degree on the space.

Since written, \eEQUAL{}\ has been verified in several
illuminating examples
\refs{\rCGOP,\rALR}.
We briefly discuss the results of
these papers and some interesting implications.
The authors of \rALR\ sought to verify \eEQUAL{}\
in a particular example through the
direct computation of both sides.
For ease of computation the authors chose to work with  the mirror pair
having $\chi = \pm 8$ in table 2. They also chose to work at large radius
on the mirror manifold (the $\chi = +8$ member of the pair) so as to
suppress the instanton contributions in \eEQUAL{}\  and leave only the
topological intersection number on the right hand side \rGP. By direct
computation, the authors of \rALR\ verified \eEQUAL{}\ in this limit.
This work also uncovered an interesting and as yet incompletely understood
subtlety: If a manifold has $h^{1,1} > 1$, one must be careful regarding
precisely how the `large radius' limit  in this multidimensional
moduli space is defined. It is possible that different limits will
pick out different theories, possibly corresponding to the inequivalent
ways in which orbifold singularities can be repaired. These ideas
and further elaborations on them are discussed in \rAL, to which
the reader is referred to for details.

The other work we would like to describe is \rCGOP. In this paper,
Candelas, de La Ossa, Green and Parkes
analyze \eEQUAL{}\ for the case of the quintic--mirror-quintic
pair \eQQ{}%
\foot{For a recent mathematical presentation of this see \rdmrsn.}.
In particular, they consider a one parameter family of
mirror manifolds given by deforming along the single K\"ahler modulus
(complex structure modulus) on the quintic (mirror quintic), as we
discussed in the general context in section 2.  Through a careful analysis
of the map between K\"ahler and complex structure deformations, the
authors are able to extract information about the right hand side
of \eEQUAL{}\ from direct calculation of the left hand side.
In particular,  they find the {\it number} of rational curves
(holomorphic instantons) of any desired degree on the quintic hypersurface.
This is to be compared with the fact that mathematicians have
to this point only succeeded in computing these numbers up to
degree three, and then only with great effort. Thus, a previously unsolved
problem in mathematics (which is quite relevant, for example, to the
manifold classification problem) is solved by mirror symmetry.

\newsec{Recent Work on Extending Mirror Symmetry}

The final topic we would like to mention is some recent work of
ours \rGPF\ which offers the possibility of extending the mirror
manifold construction reviewed in section 2 to a more general class
of theories. At the same time the work also affirms another aspect
of mirror symmetry which to this point we have not stressed. Namely,
geometrical symmetries of a given Calabi-Yau manifold manifest themselves
as {\it quantum} symmetries of the mirror manifold. Recall that
a quantum symmetry \rqs\ of a theory is one whose presence
cannot be ascertained simply by studying ordinary classical geometrical
actions. Rather, one must directly study symmetries of the full quantum
Hilbert space. For example, the $R \rightarrow 1/2R$ symmetry of string
propagation on a circle is a quantum symmetry: it certainly cannot
be ascertained from the geometrical invariances of a circle, but
nonetheless is a symmetry of the full quantum theory.

There are a number of reasons why we expect classical and quantum symmetries
to be interchanged by mirror symmetry. One concrete reason is
as follows. A conformal field theory is pieced together out of
holomorphic and antiholomorphic fields. In left-right symmetric theories,
then, if the holomorphic part of theory respects a symmetry group $G$
so will the antiholomorphic part. Thus, the full symmetry group is typically
$G \times G$. If this conformal theory admits a geometrical interpretation,
the classical geometrical symmetries are generally the diagonal or
antidiagonal subgroup of $G \times G$. This is quite familiar, for example,
from many studies of minimal model string vacua
\ref\rLR{C.A. L\"utken and G.G. Ross, Phys. Lett. {\bf 214B} (1988) 357.},
\rGP.
Now, in the case of
scaling symmetries, flipping the sign of the left moving $U(1)$ charge
interchanges the diagonal and antidiagonal subgroups of $G \times G$.
Hence, if a conformal theory admits a geometrical interpretation in terms
of mirror manifolds, classical symmetries of one manifold will be quantum
symmetries of the other and vice versa. We will see this explicitly in a
moment.

The main point of \rGPF\ is that certain apparently ill defined nonlinear
field transformations are not quite as bad as they initially seem. In fact,
they can be quite useful.  The basic idea is to introduce nonlinear field
transformations which are only one--to--one after making certain global
identifications. We aim to choose these transformations so that the latter
global identifications are in fact identical to the orbifolding
groups introduced earlier in our construction of mirror manifolds%
\foot{The reader will no doubt recall that this is essentially the program
followed in \rGVW\ in which nonlinear field transformations were used
to   pass from Landau-Ginzburg theories to Landau-Ginzburg orbifolds -- the
latter of which can be identified with Calabi-Yau manifolds.}.
This idea has also been suggested by \rCGOP\ and
\rLS. However, these field transformations do not simply provide a
new representation
for the mirror construction of \rGP, as these authors had supposed -- rather,
we  argue that
one does not generate the mirror of the {\it original} theory but
instead the mirror of a theory at a different point in the same
moduli space as the original. One part of our argument relies heavily
on the notion of quantum symmetries \rqs\ and by its success
gives us confidence that quantum symmetries of a manifold are an
accurate guide to geometrical symmetries of its mirror (as we expect
from our discussion above). This allows us to carry through
our strategy for extending the class of mirror manifolds
(as we shall see) to cover all Landau-Ginzburg theories for which
there exists a (nonlinear) field redefinition
taking the theory (up to global identifications) to Fermat form.

To illustrate these ideas, lets return to the example of the quintic
hypersurface and its mirror \eQQ{}.
Recall that the mirror manifold $M'$
is given by a $(\BZ_5)^3$
orbifold of  the Fermat quintic $M$. This
quotient may be induced by the change of variables
\eqn\ecvquin{ (X_1,X_2,X_3,X_4,X_5) = (Y_1 \, Y_3^{1/5},
                                        Y_2^{4/5} Y_5^{1/5},
                                        Y_3^{4/5} Y_4^{1/5},
                                        Y_4^{4/5} Y_2^{1/5},
                                        Y_5^{4/5})\ .}
This is because the latter change of variables is only one--to--one
if one performs a $(\BZ_5)^3$ orbifolding on the $X$ coordinates.
Notice that this transformation is singular at the origin.
In terms of these variables, the superpotential
becomes
\eqn\eWmquin{ M'' = Y_3 Y_1^5 + Y_5 Y_2^4 + Y_4 Y_3^4 + Y_2 Y_4^4 + Y_5^4\ .}
Furthermore, it is crucial to emphasize that  making \ecvquin\
one--to--one requires that we make global identifications on the $Y$ variables
as well as the $X$ variables. It is straightforward to see that
this identification on $Y$'s is a $\BZ_{256}$ action
\eqn\eIDS{(Y_1,Y_2,Y_3,Y_4,Y_5) \sim (\rho Y_1, \rho^{176} Y_2,
\rho^{251} Y_3, \rho^{20} Y_4, \rho^{64} Y_5)}
where $\rho$ is a nontrivial $256^{\rm th}$ root of unity.
These identifications are precisely implemented by embedding our equation
\eWmquin\ as
 a hypersurface $M''$ in\foot{The subscripts here denote the degree
of homogeneity of each of the homogeneous weighted projective space
coordinates.}\
\WCP{4}{41,48,51,52,64} -- the weighted projective space identifications
induce precisely \eIDS \foot{This is most easily
seen by using the generator $\lambda = \rho^{25}$ in terms of which
\eIDS\ are manifestly the weighted space identifications.}.
We see, therefore, that $M''$ is   cut out by a
polynomial of homogeneity degree 256 in this weighted projective space.
One can explicitly verify that $M'$ and $M''$ each has
a one dimensional moduli space of complex structure deformations, and 101
harmonic $(1,1)$ forms labeling K\"ahler deformations. The naive reasoning
described above might lead one to conclude the isomorphism $M' \sim M''$.

In fact, $M''$ is {\it not} the mirror of the Fermat hypersurface.
By general arguments \rqs\ the conformal field theory based on $M''$
respects a quantum $\BZ_{256}$ symmetry. The conformal field theory based on
$M$
, however,
does not respect a $\BZ_{256}$ -- classical or quantum.
 Thus, $M''$ and $M$ are not a mirror pair. As discussed in
\rGPF\ a natural conclusion to draw is that $M''$ is the mirror of a quintic
hypersurface with  a complex structure differing from the Fermat $M$. If this
is true we expect there to be a complex structure on the quintic which
gives rise to a classical $\BZ_{256}$ symmetry (which has the correct
action on the cohomology as dictated by the conformal field theory \rGPF).
Such a complex structure can be found: $M''$ is the mirror of
\eqn\eZZZquin{ \tilde M
 = X_1^5 + X_1 X_2^4 + X_2 X_3^4 + X_3 X_4^4 + X_4 X_5^4\ ,}
where the symmetry acts on the fields as
\eqn\eZZZ{ (X_1,X_2,X_3,X_4,X_5) \to
(X_1,\beta^{64} X_2, \beta^{176} X_3, \beta^{20} X_4, \beta^{251}X_5)\
}
where $\beta$ is a primitive $256^{\rm th}$ root of unity.

To understand what is going on here let us focus on the Landau-Ginzburg
theory associated with the superpotential \eWmquin.
{}From \rGVW\ we know that this theory (after the $U(1)$ projection)
describes the Calabi-Yau sigma model at a particular K\"ahler structure,
determined by the renormalization group flows.\foot{As emphasized in \rqs, this
particular structure is in fact determined by the quantum symmetry.}\
The corresponding mirror theory in quintic moduli space thus differs from the
model with which we began by an adjustment of
its {\it complex structure}  (since the  mirror of a K\"ahler
field is a complex structure field).
In this way, we wind up at \eZZZquin.
As we mentioned, the field transformation \ecvquin\ is singular. Hence,
to some extent it is surprising that this approach yields sensible results.
Undoubtedly, from the physics viewpoint, this is due to the exceptionally
well behaved operator products which arise in $N=2$ conformal theories
\rLVW; it is of interest to understand the mathematical justification
of such manipulations.

The upshot of this discussion is that we have constructed an explicit mirror
 pair
$\tilde M$ and $M''$ away from any Fermat points. By our general deformation
argument reviewed in section 2 we knew such a mirror pair existed; now we have
explicitly constructed it. For the benefit of the skeptical reader, we
hasten to point out that in \rGPF\ we verify these assertions by
deriving the differential equations satisfied by the periods, as was done
for the Fermat mirror pair in \rCGOP. We find precisely the hypergeometric
differential equation predicted by the above reasoning.
Note also that $M''$, being a hypersurface in weighted four space, is
in the list generated by \rCLS.
We now learn that this space is in fact mirror to the quintic in \CP4,
although at an unexpected point in the moduli space of the latter.
Following the same basic idea we find a number of surprising explicit
mirror pairs \rGPF. For example, preliminary work indicates that we can
build mirrors to a variety of quintic hypersurfaces by orbifolding by
a host of unexpected group actions. We list some explicit examples in
table 3.

More generally, if we are given
a Calabi-Yau theory $Q_1$ which admits a change of variables to
Fermat form (up to global identifications) then we can construct
the mirror of $Q_1$.
More precisely,
if we begin with the model $Q_1$, then this change of variables yields the
result
\eqn\egenfer{ Q_1/G_1 \sim Q_2/G_2\ ,}
where the quotient groups represent the induced identifications on both
sets of coordinates and $Q_2$ is of Fermat type, \ie\ has a point in its
moduli space corresponding to a minimal model construction. The ${}\sim{}$
in \egenfer\ implies equivalence up to deformations by twisted fields, but
for our purpose of establishing the existence of the mirror manifold, the
precise point in its moduli space is inessential. The arguments
of \rGP\ allow us to find the mirror manifold of the orbifold
appearing on the right hand side of \egenfer, expressed as a quotient of
$Q_2$ by the appropriate `dual' subgroup to $G_2$. However, we are
interested in constructing the mirror of $Q_1$, not of its quotient.
It is
here that the  work of \rGPF\ as briefly described above
gives the  solution. The left hand side of
\egenfer\ respects a quantum $G_1$ symmetry \rqs; its quotient by this is
$Q_1$. As we observed, though, on the mirror manifold of \egenfer, this
will be realized as a {\it geometrical} symmetry at some point in the
moduli space. At this point, the quotient by this symmetry will mirror the
quotient by the quantum $G_1$, yielding the mirror $Q_1 '$.

This clearly takes us a significant step beyond \rGP; an interesting open
question is to characterize precisely the set of theories for which this
procedure will generate the mirror.

\newsec{Conclusions}

In this talk we have reviewed the initial speculations
\refs{\rDIXON,\rLVW}\
and subsequent
explicit demonstration \rGP\
of the existence of mirror symmetry -- a symmetry
whereby two Calabi-Yau manifolds whose Hodge diamonds differ by a ninety-degree
rotation are shown to correspond to the same conformal field theory. We
have stressed that this symmetry interchanges the roles of complex structure
and K\"ahler moduli on the two manifolds and, correspondingly, interchanges
classical and quantum symmetries. We have also emphasized one especially
important implication of mirror symmetry initially found in \rGP --
namely eqn. \eEQUAL{}.
This is a rather remarkable equality in that an infinite instanton sum
is reexpressed as a simple geometrical integral.
Although \eEQUAL{}\ follows directly from the general arguments of \rGP,
it is most pleasing that it has been directly verified in a special case
\rALR. Furthermore, the power of \eEQUAL{}, as briefly reviewed, has been
recently exploited by \rCGOP\ to solve a previously unresolved mathematical
issue: the determination of the number of rational curves of arbitrary
degree on the quintic threefold.

There are a number of compelling unresolved issues surrounding mirror
symmetry. The two most prominent are: 1) Is mirror symmetry fully
general in the sense that any Calabi-Yau $n$--fold has a mirror
manifold whose Hodge diamond is rotated by ninety-degrees and which
corresponds to the same conformal field theory? The speculative
discussions based on naturality certainly seem to support this statement.
At this time, as discussed, mirror symmetry has only been established
for a particular subclass of Calabi-Yau manifolds.
2) What is the underlying mathematical reason for the existence of mirror
manifolds? Associated with this is the question of the role played
by our orbifolding prescription -- is it fundamental to the existence
of mirror manifolds or simply a useful tool which is central to the
construction of all presently known examples?
Such questions are under active study; with some luck insight into
their answers will not be long in coming.

\medskip

We thank N. Elkies, K. Intriligator, and C. Vafa for discussions.
B.R.G.~ is supported by NSF grant PHY-87-15272,
and M.R.P.~ by DOE grant DE-AC-02-76ERO3075.

\vfill \eject
\vfill

\moveleft .7 truein
\vbox{
$$\vbox{\offinterlineskip
\hrule height 1.1pt
\halign{&\vrule width 1.1pt#&\strut\quad\hfil#\hfil\quad&
\vrule#&\strut\quad\hfil#\hfil\quad&
\vrule#&\strut\quad\hfil#\hfil\quad&
\vrule#&\strut\quad\hfil#\hfil\quad&
\vrule#&\strut\quad\hfil#\hfil\quad&\vrule width 1.1pt#\cr
height3pt&\omit&&\omit&&\omit&&\omit&&\omit&\cr
&$M$&&$h^{2,1}$&&$h^{1,1}$&&$S$&&$M'$&\cr
height3pt&\omit&&\omit&&\omit&&\omit&&\omit&\cr
\noalign{\hrule}
height3pt&\omit&&\omit&&\omit&&\omit&&\omit&\cr
&$z_1^5 + z_2^5 + z_3^5 + z_4^5 + z_5^5 = 0 $&
&101&&1&
&$\BZ_5^5$&&$M/ ( \BZ_5^3 ) $&\cr
height3pt&\omit&&\omit&&\omit&&\omit&&\omit&\cr
\noalign{\hrule}
height3pt&\omit&&\omit&&\omit&&\omit&&\omit&\cr
&$z_1^5 + z_2^{10} + z_3^{10} + z_4^{10} + z_5^2 = 0 $&
&145&&1&
&$\BZ_5 \times \BZ_{10}^3\times \BZ_2$&&$M/ ( \BZ_{10}^2 ) $&\cr
height3pt&\omit&&\omit&&\omit&&\omit&&\omit&\cr
\noalign{\hrule}
height3pt&\omit&&\omit&&\omit&&\omit&&\omit&\cr
&${\textstyle z_1^3 + z_2^3 + z_3^3 + z_4^3 = 0}
\atop {\textstyle z_1 x_1^3 + z_2 x_2^3 + z_3 x_3^3 = 0} $&
&35&&8&
&$\BZ_3 \times \BZ_9^3$&&$M/ ( \BZ_3\times \BZ_9 ) $&\cr
height3pt&\omit&&\omit&&\omit&&\omit&&\omit&\cr
\noalign{\hrule}
height3pt&\omit&&\omit&&\omit&&\omit&&\omit&\cr
&$z_1^3 + z_2^ 4 + z_3^5 + z_4^6 + z_5^{20} = 0 $&
&47&&23&
&$\BZ_3 \times \BZ_4\times \BZ_6\times \BZ_{20}$&&$M/ ( \BZ_2 ) $&\cr
height3pt&\omit&&\omit&&\omit&&\omit&&\omit&\cr
\noalign{\hrule}
height3pt&\omit&&\omit&&\omit&&\omit&&\omit&\cr
&$z_1^3 + z_2^4 + z_3^4 + z_4^{12} + z_5^{12} = 0 $&
&89&&5&
&$\BZ_3\times \BZ_4^2\times \BZ_{12}^2$&&$M/ ( \BZ_3\times \BZ_4^2 ) $&\cr
height3pt&\omit&&\omit&&\omit&&\omit&&\omit&\cr
\noalign{\hrule}
height3pt&\omit&&\omit&&\omit&&\omit&&\omit&\cr
&$z_1^4 + z_2^6 + z_3^{21} + z_4^{28} + z_5^2 = 0 $&
&52&&28&
&$\BZ_4\times \BZ_6\times \BZ_{21}\times \BZ_{28}$&&$M/ ( \BZ_2 ) $&\cr
height3pt&\omit&&\omit&&\omit&&\omit&&\omit&\cr
\noalign{\hrule}
height3pt&\omit&&\omit&&\omit&&\omit&&\omit&\cr
&$z_1^5 + z_2^6 + z_3^{10} + z_4^{30} + z_5^2 = 0 $&
&91&&7&
&$\BZ_5\times \BZ_6\times \BZ_{10}\times \BZ_{30}$&
&$M/ ( \BZ_{2}\times \BZ_{10} ) $&\cr
height3pt&\omit&&\omit&&\omit&&\omit&&\omit&\cr}
\hrule height 1.1pt}$$}
\vskip.3in
\centerline{Table 1}
\centerline{Mirrors $M'$ of Some Minimal Model Compactifications $M$}
\vskip .3in
The equations listed define $M$ as a hypersurface in an appropriate weighted
projective 4-space (such that the equation is homogeneous). The last column
gives the mirror manifold $M'$ as a quotient of $M$. The third entry
defines a submanifold of $\CP3 \times \CP2$, demonstrating that the restriction
to minimal models sometimes extends beyond hypersurfaces.

\vfill \eject
\vfill

$$\vbox{\offinterlineskip
\hrule height 1.1pt
\halign{&\vrule width 1.1pt#&\strut\quad\hfil#\hfil\quad&
\vrule#&\strut\quad\hfil#\hfil\quad&
\vrule#&\strut\quad\hfil#\hfil\quad&
\vrule#&\strut\quad\hfil#\hfil\quad&\vrule width 1.1pt#\cr
height3pt&\omit&&\omit&&\omit&&\omit&\cr
&Symmetries&&$h^{2,1}$&&$h^{1,1}$&&$\chi$&\cr
height3pt&\omit&&\omit&&\omit&&\omit&\cr
\noalign{\hrule}
height3pt&\omit&&\omit&&\omit&&\omit&\cr
&&&101&&1&&-200&\cr
height3pt&\omit&&\omit&&\omit&&\omit&\cr
\noalign{\hrule}
height3pt&\omit&&\omit&&\omit&&\omit&\cr
&[0,0,0,1,4]&&49&&5&&-88&\cr
height3pt&\omit&&\omit&&\omit&&\omit&\cr
\noalign{\hrule}
height3pt&\omit&&\omit&&\omit&&\omit&\cr
&[0,1,2,3,4]&&21&&1&&-40&\cr
height3pt&\omit&&\omit&&\omit&&\omit&\cr
\noalign{\hrule}
height3pt&\omit&&\omit&&\omit&&\omit&\cr
&${\textstyle [0,1,1,4,4]} \atop
{\textstyle [0,1,2,3,4]} $&&21&&17&&-8&\cr
height3pt&\omit&&\omit&&\omit&&\omit&\cr
\noalign{\hrule}
height3pt&\omit&&\omit&&\omit&&\omit&\cr
&[0,1,1,4,4]&&17&&21&&8&\cr
height3pt&\omit&&\omit&&\omit&&\omit&\cr
\noalign{\hrule}
height3pt&\omit&&\omit&&\omit&&\omit&\cr
&${\textstyle [0,1,3,1,0]} \atop
{\textstyle [0,1,1,0,3]} $&&1&&21&&40&\cr
height3pt&\omit&&\omit&&\omit&&\omit&\cr
\noalign{\hrule}
height3pt&\omit&&\omit&&\omit&&\omit&\cr
&$ {\textstyle [0,1,4,0,0]} \atop
{\textstyle [0,3,0,1,1]} $&&5&&49&&88&\cr
height3pt&\omit&&\omit&&\omit&&\omit&\cr
\noalign{\hrule}
&[0,1,0,0,4]&&&&&&&\cr
&[0,0,1,0,4]&&1&&101&&200&\cr
&[0,0,0,1,4]&&&&&&&\cr
height3pt&\omit&&\omit&&\omit&&\omit&\cr}
\hrule height 1.1pt}$$
\vskip.3in
\centerline{Table 2}
\centerline{Orbifolds of $z_1^5 + z_2^5 + z_3^5 + z_4^5 + z_5^5 = 0$ in \CP4}
\vskip .3in
The generators of each symmetry group are represented by the powers of a
fundamental fifth root of unity by which they multiply the homogeneous
coordinates of $\CP4$. Thus $[0,0,0,1,4]$, for example, represents
$$(z_1,z_2,z_3,z_4,z_5) \to (z_1,z_2,z_3,\alpha z_4,\alpha^4 z_5)$$
with $\alpha^5 = 1$.

\vfill \eject
\vfill

$$\vbox{\offinterlineskip
\hrule height 1.1pt
\halign{&\vrule width 1.1pt#&\strut\quad\hfil#\hfil\quad&
\vrule#&\strut\quad\hfil#\hfil\quad&
\vrule#&\strut\quad\hfil#\hfil\quad&\vrule width 1.1pt#\cr
height3pt&\omit&&\omit&&\omit&\cr
&$M$&&$M'$&&$G$ Generators&\cr
height3pt&\omit&&\omit&&\omit&\cr
\noalign{\hrule}
height3pt&\omit&&\omit&&\omit&\cr
&&&&&[0,0,0,1,4]&\cr
&$z_1^5 + z_2^5 + z_3^5 + z_4^5 + z_5^5 = 0$&
&$M/ ( \BZ_5^3 ) $&
&[0,0,1,0,4]&\cr
&&&&&[0,1,0,0,4]&\cr
height3pt&\omit&&\omit&&\omit&\cr
\noalign{\hrule}
height3pt&\omit&&\omit&&\omit&\cr
&$z_1^4 z_2 + z_2^4 z_3 + z_3^4 z_4 + z_4^4 z_5 + z_5^4 z_1 = 0$&
&$M/ ( \BZ_{41} ) $&
&[1,37,16,18,10]&\cr
height3pt&\omit&&\omit&&\omit&\cr
\noalign{\hrule}
height3pt&\omit&&\omit&&\omit&\cr
&$z_1^4 z_2 + z_2^4 z_3 + z_3^4 z_4 + z_4^4 z_1 + z_5^5 = 0$&
&$M/ ( \BZ_{51} ) $&
&[1,37,16,38,0]&\cr
height3pt&\omit&&\omit&&\omit&\cr
\noalign{\hrule}
height3pt&\omit&&\omit&&\omit&\cr
&$z_1^4 z_2 + z_2^4 z_3 + z_3^4 z_1 + z_4^5 + z_5^5 = 0$&
&$M/ ( \BZ_5 \times \BZ_{13} ) $&
&${\textstyle [0,0,0,1,4]} \atop
{\textstyle [1,9,3,0,0]} $&\cr
height3pt&\omit&&\omit&&\omit&\cr
\noalign{\hrule}
height3pt&\omit&&\omit&&\omit&\cr
&&&&&[0,0,1,0,4]&\cr
&$z_1^4 z_2 + z_2^4 z_1 + z_3^5 + z_4^5 + z_5^5 = 0$&
&$M/ ( \BZ_5^2 \times \BZ_3 ) $&
&[0,0,0,1,4]&\cr
&&&&&[1,2,0,0,0]&\cr
height3pt&\omit&&\omit&&\omit&\cr
\noalign{\hrule}
height3pt&\omit&&\omit&&\omit&\cr
&$z_1^4 z_2 + z_2^4 z_1 + z_3^4 z_4 + z_4^4 z_5 + z_5^4 z_3 = 0$&
&$M/ ( \BZ_3 \times \BZ_{13} ) $&
&$ {\textstyle [1,2,0,0,0]} \atop
{\textstyle [0,0,1,9,3]} $&\cr
height3pt&\omit&&\omit&&\omit&\cr}
\hrule height 1.1pt}$$
\vskip.3in
\centerline{Table 3}
\centerline{Quintic $M$ Hypersurfaces and Their Mirrors $M'$}
\vskip.3in
The indicated quotient of the manifold $M$ is its mirror
$M'$. The group actions are denoted by the powers of an appropriate primitive
root of unity by which they multiply the coordinates on \CP4, as in table 2.

\vfill \eject

\listrefs

\bye